\documentclass[sigconf]{acmart}
\AtBeginDocument{%
  }

\setcopyright{acmlicensed}

\copyrightyear{2025} 
\acmYear{2025} 
\setcopyright{cc}
\setcctype{by}
\acmConference[CCS '25]{Proceedings of the 2025 ACM SIGSAC Conference on Computer and Communications Security}{October 13--17, 2025}{Taipei, Taiwan}
\acmBooktitle{Proceedings of the 2025 ACM SIGSAC Conference on Computer and Communications Security (CCS '25), October 13--17, 2025, Taipei, Taiwan}\acmDOI{10.1145/3719027.3765221}
\acmISBN{979-8-4007-1525-9/2025/10}

\usepackage{booktabs}
\usepackage[table]{xcolor}
\usepackage{tabularx}
\usepackage{longtable}
\usepackage{adjustbox}
\usepackage{afterpage}

\usepackage{amsthm}

\newcommand{\boldparagraph}[1]{\paragraph{#1}}

\makeatletter
\renewcommand\paragraph{\@startsection{paragraph}{4}{0\parindent}%
    {0.4ex plus 0.8ex minus 0.2ex}%
    {0ex}%
    {\normalfont\normalsize\bfseries\maybe@addperiod}
}
\newcommand{\maybe@addperiod}[1]{%
    #1\@addpunct{.}\enspace%
}
\makeatother

\usepackage{xargs}

\newcommand\revision[1]{{{\color{black}#1}}}

\newcommand\rev[1]{{{\color{black}#1}}}

\usepackage[style=american,threshold=2]{csquotes}

\usepackage[usetwoside=true]{mdframed}
\usepackage{tcolorbox}
\tcbuselibrary{breakable}
\tcbuselibrary{skins}

\definecolor{darkgray}{gray}{0.3}

\tcbset{}
\newtcolorbox{summaryBox}[2][]
{
    enhanced,
    breakable,
    frame hidden,
    fontupper       = \small,
    fontlower       = \footnotesize,
    borderline west = {1.5pt}{0pt}{black},
    colback         = white,
    size            = fbox,
    coltitle        = black,
    title           = {\hspace{0.5em}\color{black}#2. },
    attach title to upper,
    #1,
}

\begin{document}


\title{Extended Version: Security and Privacy Perceptions of Pakistani Facebook Matrimony Group Users}

\thanks{\textcolor{red}{\textbf{A shortened version of this paper appears in the Proceedings of the 2025 ACM SIGSAC Conference on Computer and Communications Security (ACM CCS 2025), Taipei, Taiwan. This is the full version with the interview guide, screening survey, Facebook post mockups, and codebook in the appendix. All additional study materials are available at {\url{https://doi.org/10.5281/zenodo.17020107}}.}}}


\author{Mah Jan Dorazahi}
\authornote{Both authors contributed equally to this research.}
\affiliation{
  \institution{Paderborn University}
  \city{Paderborn}
  \country{Germany}
}
\email{mahieg.joie@gmail.com}

\author{Deepthi Mungara}
\authornotemark[1]
\affiliation{
  \institution{Paderborn University}
  \city{Paderborn}
  \country{Germany}
}
\email{deepthi.mungara@uni-paderborn.de}

\author{Yasemin Acar}
\affiliation{
  \institution{Paderborn University \& The George Washington University}
  \city{Paderborn}
  \country{Germany}
}
\email{yasemin.acar@uni-paderborn.de}

\author{Harshini Sri Ramulu}
\affiliation{
  \institution{Paderborn University}
  \city{Paderborn}
  \country{Germany}
}
\email{harshini.sri.ramulu@uni-paderborn.de}









\renewcommand{\shortauthors}{Mah Jan Dorazahi and Deepthi Mungara et al.}

\begin{abstract}
In Pakistan, where dating apps are subject to censorship, Facebook matrimony groups---also referred to as \textit{marriage groups}---serve as alternative virtual spaces for members to search for potential life partners. To participate in these groups, members often share sensitive personal information such as photos, addresses, and phone numbers, which exposes them to risks such as fraud, blackmail, and identity theft.
To better protect users of Facebook matrimony groups, we need to understand aspects related to user \textit{safety}, such as how users perceive risks, what influences their trust in sharing personal information, and how they navigate security and privacy concerns when seeking potential partners online.
In this study, through 23 semi-structured interviews, we explore how Pakistani users of Facebook matrimony groups perceive and navigate risks of sharing personal information, and how cultural norms and expectations influence their behavior in these groups.

We find elevated privacy concerns among participants, leading them to share limited personal information and creating mistrust among potential partners. Many also expressed concerns about the authenticity of profiles and major security risks, such as identity theft, harassment, and social judgment. Our work highlights the challenges of safely navigating Facebook matrimony groups in Pakistan and offers recommendations for such as implementing stronger identity verification by group admins, enforcing stricter cybersecurity laws, clear platform guidelines to ensure accountability, and technical feature enhancements---including restricting screenshots, picture downloads, and implementing anonymous chats---to protect user data and build trust.
\end{abstract}

\ccsdesc[500]{Security and Privacy~Human and societal aspects of security and privacy}

\keywords{facebook matrimony groups, privacy, security, pakistani users}

\maketitle

\section{Introduction}
\label{sec:intro}
In Pakistan, matrimonial groups and websites are often used by individuals seeking potential long-term romantic partners~\cite{wollburg2016history}. \rev{Traditional matrimonial practices solely existed offline, where matchmaking was \textit{arranged} through families and trusted social circles. Owing to digitization, these practices have shifted online to Facebook groups such as Soul Wonders, Two Rings, and Soul Sisters,\footnote{\rev{Examples of popular Facebook matrimony groups: \href{https://www.facebook.com/groups/905284146926802}{Soul Wonders}, \href{https://www.facebook.com/groups/3828049250554649}{Two Rings}, and \href{https://www.facebook.com/groups/3828049250554649}{Soul Sisters}.}} and to dedicated matrimonial websites like Pure Matrimony and Shadi.com.\footnote{\rev{Popular matrimonial websites used in Pakistan: \href{https://purematrimony.com/}{https://purematrimony.com/} and \href{https://www.shadi.com/}{https://www.shadi.com/}.}} In early 2020, major dating apps like Tinder were banned in Pakistan, leading to the rise of culturally sensitive platforms like Muzz.\footnote{\rev{Pakistani dating app \href{https://www.muzz.com/}{https://www.muzz.com/}---formerly Muzzmatch---used as an alternative to \href{https://www.tinder.com/}{Tinder}.}}}
While these platforms try to align with cultural practices of long-term relationships in Muslim communities, they face resistance from traditional families and skepticism about their ability to help young people---especially women---make informed marriage decisions without traditional support networks~\cite{marketing2024}. Therefore, Facebook remains a popular platform among families for matchmaking~\cite{tayyab2022pak, pakistanismatchmaking}. \revision{Further, prior work indicates that women feel comfortable expressing themselves on Facebook, overcoming challenges imposed by cultural norms, such as discussing taboo topics pertaining to relationships and sex, and interacting with people of different genders~\cite{ammari2022moderation}.}
~\revision{Facebook groups serve a multifaceted purpose---allowing families to take control in seeking partners and also providing a space for women and marginalized groups to explore relationships---and remain widely popular due to their usability, affordability, and accessibility~\cite{statcounter,abbas2020effect, krishnan2024study}.} 
While these groups present opportunities to a bigger pool of potential partners, they also uphold cultural values by allowing families to be more involved in matchmaking~\cite{rguibi2023towards}. However, users are exposed to scams~\cite{whitty2015anatomy}, doxxing~\cite{eckert2020doxxing}, stalking~\cite{chugh2022stalking}, and privacy violations~\cite{sarikakis2017social}. These risks are often interwoven with gender and cultural norms, where women are monitored and scrutinized more, since their actions are closely tied to the notion of family honor~\cite{irfan2009honor,mahmood2024women, afnan2022aunties}. Even though these issues are prevalent, security and privacy concerns in Facebook matrimony groups remain understudied.
\revision{Our study addresses this gap by examining how Pakistani Facebook matrimony group users navigate security, privacy, and trust, drawing key insights from participants' lived experiences, including personal practices and community-level observations.}

Through 23 semi-structured interviews with Facebook matrimony group users, we explore participants' security and privacy concerns, experiences of scams, frauds, and perceived threats. We also examined factors that influence trust and mistrust in potential matches, and the reasons for sharing and withholding personal information. Specifically, we aim to address the following four research questions:

\noindent\textbf{RQ1:} \textit{What information do Pakistani users share on their profiles in Facebook matrimony groups? What factors influence them to share or withhold information?} \\
\textbf{RQ2:}  \textit{What are their perceptions of security and privacy risks, concerns and threats of using Facebook matrimony groups?} \\
\textbf{RQ3:} \textit{What factors influence users to trust the authenticity of matrimonial profiles they encounter?} \\
\textbf{RQ4:} \textit{How are users' security, privacy, and safety currently supported by platforms and group administrators, and how can this be improved?} 

We find that security and privacy concerns strongly influence participants' trust and self-disclosure behavior in matrimonial groups. To protect themselves from matrimony scams, identity theft, cyber threats, and social judgment, participants adopted various strategies such as verifying identities of matches across platforms (e.g., on LinkedIn and Instagram), avoiding sharing financial information, and using progressive disclosure. We recommend implementing a robust platform-level verification process, increasing users' awareness for safely sharing personal information, and developing culturally appropriate solutions to mitigate associated privacy risks.

\rev{\section{Background and Related Work}}
\label{sec:background and relwork}
In this section, we present related work on privacy and trust in social media and dating platforms used for finding romantic relationships. \rev{We also provide a brief background on the influence of cultural factors in online dating and matchmaking in Pakistan, and the} impact of cybersecurity laws in Pakistan. To illustrate these practices in Pakistan, Figure~\ref{fig:man} shows a mock-up of a typical matrimonial post, reflecting the cultural norm of sharing personal details and expectations of a future partner.

\afterpage{
\begin{figure}[t]
\begin{center}

\includegraphics[width=\columnwidth]{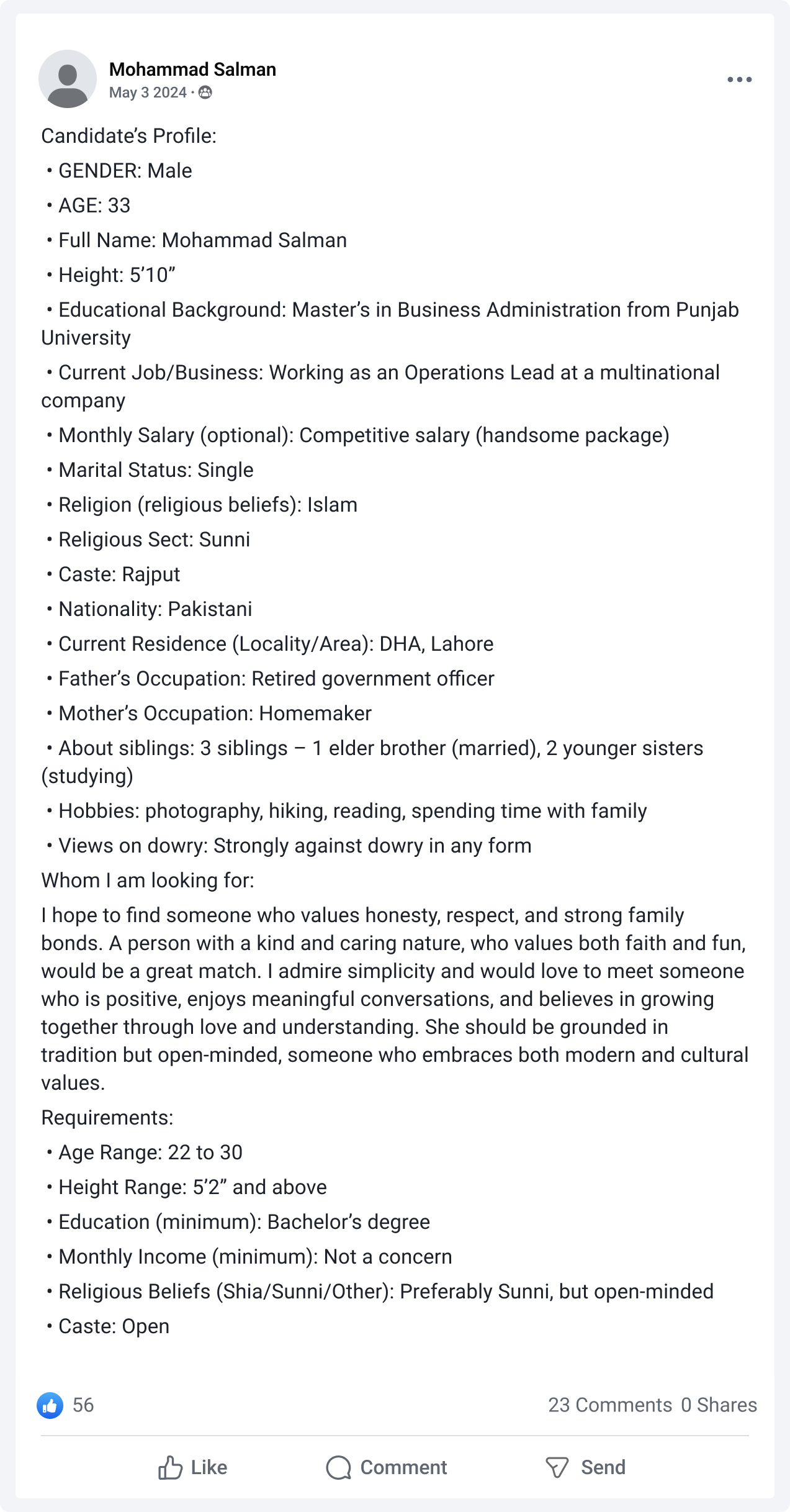}
\caption{A mock-up of an example Facebook post containing all the information shared on matrimonial profiles.}
\label{fig:man}
\end{center}
\end{figure}}

\subsection{Privacy and trust in social media \rev{and dating platforms}} \label{subsec:global_dating}
\revision{Privacy and trust in social media is a widely studied topic in prior work, and several studies have explored how perceptions of security and privacy in a social media platform affect trust and information sharing behaviors~\cite{acquisti2006imagined, dhami2013impact}. Trust is affected by the design of the social media platform, where users have the ability to control privacy settings and decide who can view their information~\cite{al2024building}. 
User characteristics, such as gender and educational background influence mistrust in online dating, often relating to concerns about fake profiles, identity theft, and social engineering~\cite{kamble_issues_2019, whitty2012online}.}  \rev{Online dating platforms and apps introduce risks such as romance scams~\cite{buchanan2014online}, identity theft~\cite{wani2017sneak}, social engineering~\cite{mouton2014social}, and cyberbullying~\cite{muttaqin2020cyberbullying, orlu2017discursive}. These risks often disproportionately harm women, reinforcing gendered oppression in online dating spaces~\cite{orlu2017discursive}}. 
To address concerns of trust in matrimonial contexts, users try to verify potential matches themselves by scrutinizing their posts and online behavior, and using one-on-one conversations to gauge authenticity of a selected profile~\cite{al2013males}.
\rev{Users generally are required to share personal information to find matches~\cite{graciyal2020virtual, cobb2017public}. The amount of personal information users share on their matrimony profiles are closely tied to their seriousness in finding a match, highlighting a common belief that more openness attracts more matches, often leaving users in a dilemma of whether or not to share information~\cite{graciyal2020virtual}.} Additionally, users make privacy and information sharing decisions based on factors like their ability to control the amount of information they share, the size of the community, their interest in the community, and visibility of their profile~\cite{acquisti2006imagined, syn2015social}. 
In Facebook matrimony groups, users share their profile and sensitive personal information publicly with potential matches, thus navigating a complex trade-off between personal privacy and information sharing. \rev{Therefore, in our study, we explore how Facebook matrimony group users navigate trust and safety in matchmaking communities, how they make trust-based decisions, and what influences their willingness to share personal information.}

\subsection{Cultural contexts and social media dating and matchmaking in Pakistan} \label{subsec:pakistan dating}
Facebook matrimony groups in Pakistan offer a virtual space to connect with potential partners by bypassing both banned apps and traditional matchmaking methods~\cite{gillani2024usage, rguibi2023towards}. Prior work highlights both the success and challenges of Facebook matrimony groups; many users have successfully found long-term partners, but they also face challenges like inefficient profile filtering and low response rates from potential matches~\cite{gillani2024usage}.
\rev{Users often use Facebook also as a tool for investigating and assessing credibility of potential partners by scrutinizing profile details, posts, and friends lists~\cite{facebook_conflicts_2018}.}

Cultural factors, such as~\rev{family involvement and gender}, also shape how these platforms are used, for instance, allowing family members to post on behalf of match-seekers~\cite{al2017against}. Contrarily, some users prefer to not disclose their dating habits to families, and deliberately hide any related social media activity from their families. This selective sharing behavior is referred to as ``family anonymity''~\cite{al2024building, roy2020fake}. Further, in some patriarchal societies in Pakistan, women are particularly vulnerable to targeted online abuse; they have restricted access to digital media and phones, and consequently rely on male relatives to make decisions about security and information sharing, especially in matrimonial groups~\cite{khan2022cyber, naveed2022ask, sambasivan2018privacy}. To amplify this, prior research shows dating apps can facilitate gendered abuse, highlighting the complex dynamics between gender, digital access, and autonomy~\cite{gillett2023not}. In our study, we specifically examine \rev{how cultural norms shape privacy expectations and influence information-sharing behaviors in Facebook matrimony groups. We draw on cultural contexts like gender influences and family involvement in developing our interview guide  (Appendix~\ref{tab:interview-guide}) and understanding cultural nuances during data analysis.} 

\subsection{Legal cybersecurity landscape in Pakistan} \label{subsec:sec_laws}
Several studies have explored the legal landscape of cybercrimes in Pakistan, finding a prevalence of blackmail and impersonation ~\cite{akhlaq2021cybercrime}, following increasing digitization of  services~\cite{zahoor2020cyber}. 
Prior work has investigated the challenges in enforcing cyber security laws, identifying corruption among other factors inhibiting their implementation~\cite{bokhari2023quantitative}. Additionally arguments prevail that existing laws do not protect individuals, institutions, or the state in cyberspace, and require updates and legal clarity~\cite{khan2020cybersecurity}. Similarly, prior work highlights the shortcomings of the Prevention of Electronic Crimes Act (PECA) 2016 in addressing modern cyber threats in Pakistan, stressing the need for better enforcement, stakeholder collaboration, and legislative reforms to tackle identity theft and online harassment~\cite{zahid2024cybercrime}. Following the presence (or absence) of sufficient legal protections for users from online harms, we investigate users' perceptions of legal support in navigating online challenges related to privacy invasions in Facebook matrimony groups.

\section{Methodology}
\label{sec:methodology}
We conducted 23 semi-structured interviews to understand security and privacy perceptions, and risks associated with using Facebook matrimony groups. In this section, we provide an overview of our recruitment process, data collection, analysis, ethical considerations, and limitations.

\subsection{Study procedure}
\boldparagraph{Participant recruitment} We recruited seven participants through personal connections; four were distant contacts with whom the researchers had minimal or no contact. The remaining 16 participants were recruited through Facebook matrimony groups. Our selection criteria included one or all of the following: (a) being a user of at least one Facebook matrimony group, (b) posting a matrimony profile for themselves or for someone else, and (c) being an admin of a Facebook matrimony group. 

After we exhausted personal contacts, we decided to post in matrimony groups to recruit more users. Most of these groups are controlled by admins; therefore, we reached out to them for permission to post recruitment messages in the group. We contacted seven admins, and only two responded, allowing us to recruit from these groups. We also reached out to one group that was not a dedicated matrimony group but had matrimony posts. All of these groups are run by and for people from Pakistan and the Pakistani diaspora. One author, who is a Facebook user, reached out to admins and posted recruitment messages from their personal Facebook profile. In the recruitment message, we posted a link to our recruitment survey (Appendix~\ref{tab:Participant-recruitment-survey}), and aimed for a diverse sample in terms of marital status, gender, current location, and educational backgrounds. Our demographic table is available in Table~\ref{table:demographics}, and all the study materials are available at {\url{https://doi.org/10.5281/zenodo.17020107}}.

\begin{table}[!ht]
\centering
\begin{tabular}{@{}llllll@{}} 
\toprule 
\multicolumn{2}{c}{\textbf{Posted for}} & \multicolumn{2}{c}{\textbf{Age}} & \multicolumn{2}{c}{\textbf{Gender}} \\
\cmidrule(r){1-2} \cmidrule(lr){3-4} \cmidrule(l){5-6} 
Myself         & 4  & 18--24 & 5  & Man   & 8  \\
Someone else   & 6  & 25--34 & 16 & Woman & 15 \\
Not posted (but uses) & 12 & 35--44 & 1  &       &    \\
Prefer not     & 1  & 45--55 & 1  &       &    \\
\midrule
\multicolumn{2}{c}{\textbf{Education}} & \multicolumn{2}{c}{\textbf{Degree in CS}} & \multicolumn{2}{c}{\textbf{Employment}} \\
\cmidrule(r){1-2} \cmidrule(lr){3-4} \cmidrule(l){5-6} 
Secondary  & 2  & Yes & 7  & Full-time   & 14 \\
Bachelor   & 12 & No  & 16 & Part-time   & 4  \\
Master     & 8  &     &    & Not working & 4  \\
PhD        & 1  &     &    & Other       & 1  \\
\midrule
\multicolumn{2}{c}{\textbf{Marital Status}} & \multicolumn{2}{c}{\textbf{Residence}} & \multicolumn{2}{c}{\textbf{Total}} \\
\cmidrule(r){1-2} \cmidrule(lr){3-4} \cmidrule(l){5-6} 
Single     & 12 & Pakistan & 13 & Participants & 23 \\
Married    & 9  & Germany  & 5  &              &    \\
Divorced   & 2  & UK       & 4  &              &    \\
           &    & Canada   & 1  &              &    \\
\bottomrule
\end{tabular}
\caption{Interview participant demographics.}
\label{table:demographics}
\end{table}

\boldparagraph{Interview guide creation}
The interview guide was structured around research questions that explore user perceptions related to information sharing in Facebook matrimony groups. We developed our interview guide based on gaps we identified in prior work regarding how users navigate privacy, trust, and safety in informal online matchmaking spaces (Section~\ref{subsec:pakistan dating}). In our interview guide, we specifically focused on capturing users' perspectives on sharing personal information and the role of trust. Furthermore, we focused on reasons for sharing or withholding certain personal details, safety measures users adopt, perceived threats associated with the information disclosure, how trust influences self-disclosure and profile verification, and the support received from group admins and the platform in facilitating a safe and secure matchmaking process. Our interview guide can be found in Appendix~\ref{tab:interview-guide}.

\boldparagraph{Interview procedure}
We conducted four pilot interviews (two in English and two in Urdu) to test the comprehensibility and the flow of the interview questions. Based on the feedback from the pilots, we iterated the interview guide: we added new questions, rephrased some, and removed some questions. For the actual interviews, we conducted in-depth interviews with 23 participants; all interviews were conducted in English or Urdu by the lead author, who was able to speak both languages. Interviews lasted 25 minutes on average. We used our university Zoom to conduct all interviews; we audio recorded all interviews and used Zoom transcriptions for all English interviews. For Urdu transcripts, the lead author manually transcribed four interviews using audio recordings, which was manually intensive. For the next six interviews, we used MacWhisper~\cite{macwhisper2024} to speed up the transcription process. All interview transcripts were carefully de-identified and stored on our self-hosted secure cloud service. 

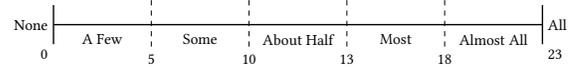
\begin{figure}[ht]
\centering
\scalebox{0.65}{
\begin{tikzpicture}
    \draw[thick] (0,0) -- (10,0);
    \draw[thick] (0,0.4) -- (0,-0.4);
    \draw[thick] (10,0.4) -- (10,-0.4);
    \node[left] at (0,0) {None};
    \node[left] at (0,-0.6) {0};
    \node[right] at (10,0) {All};
    \node[right] at (10,-0.6) {23};
    \foreach \x in {2,4,6,8} {
        \draw[dashed] (\x,0.5) -- (\x,-0.5);
    }
    \node at (1,-0.3) {A Few};
    \node at (3,-0.3) {Some};
    \node at (5,-0.3) {About Half};
    \node at (7,-0.3) {Most};
    \node at (9,-0.3) {Almost All};
    \node at (2,-0.7) {5};
    \node at (4,-0.7) {10};
    \node at (6,-0.7) {13};
    \node at (8,-0.7) {18};
\end{tikzpicture}
}
\caption{Quantifier scale with corresponding participant counts used throughout the results.}
\label{fig:quantifier-scale}
\end{figure}

\boldparagraph{Data analysis}
We used qualitative open coding to analyze the interview data~\cite{williams2019art}. We built our codebook iteratively; initially, one researcher---who also coded all the interviews---independently coded three transcripts to create an initial codebook. A second researcher used this codebook to independently code two transcripts; the two researchers met and discussed disagreements and adjusted the codebook. Next, the researchers coded three additional transcripts independently and reached a poor inter-coder agreement value, after which they resolved disagreements and iterated the codebook. This process continued until the researchers reached a value of $\kappa$ $>$ 0.82~\cite{kuckartz2019analyzing}, resulting in the secondary coder coding 30\% of the transcripts. We then used an affinity diagram to organize our codes into meaningful categories to inform our results section. \rev{To convey a general idea of the weight of our findings, we use a quantifier scale (see Figure~\ref{fig:quantifier-scale}), which we refer to throughout our results. We use this scale for consistency and to avoid using numerical counts and percentages in presenting our qualitative findings.}

\subsection{Ethical considerations and reflexivity}
We did not obtain an IRB approval for this study because our institute did not require ethical review; however, we followed the ethical guidelines in the Menlo Report and data protection guidelines throughout the research process~\cite{bailey2012menlo}. We added an informed consent form to the recruitment survey, and participants were only able to fill the survey, after reading the consent form. Further, we obtained verbal consent for recording and transcribing interview data at the start of the interview. We carefully de-identified any personally identifying information and only stored de-identified transcripts. 

Due to the sensitive nature of our study, we placed emphasis on participants' right to withdraw from the study at any point without bearing any consequences. We provided an option for participants to not answer questions they were uncomfortable with. \rev{We conducted interviews in Urdu, English, or a combination of both, based on participants' preferences. Although the interviewer was fluent in both languages, certain concepts lack direct translations to English, leading to loss of nuance of certain topics during translation, both in the interviews and analysis.} Therefore, we encouraged participants to ask questions and allowed them to seek clarifications when needed, and we also offered to switch to Urdu if there were any issues. We re-emphasized, when recalling experiences with scams and privacy violations, that participants could refrain from answering these questions. \revision{We were unable to compensate participants because our institute did not have feasible mechanisms to compensate in a foreign currency.} In our data analysis, we were mindful of the language used by our participants that reflected colorism, sexism, ageism (particularly toward women over the age of 30), and stigma toward divorced people. In the presentation of this paper, we paraphrase any occurrences of these terminologies if used in direct quotes or just opt for summarizing quotes where necessary. We do not reflect any thoughts of the participants and do not endorse any of these stigmatizing values ourselves. 

\subsection{Limitations}
There are many limitations to our study, which includes the exclusive focus on Facebook as a matrimonial platform, limiting the generalizability of our findings to other online services like matrimonial websites and applications. With the qualitative nature of the study, our findings may be subject to participant self-censorship, observer bias, and recall bias~\cite{bradburn1987answering,olson2014ways}. \revision{Addtionally, we interviewed seven participants from personal contacts; their responses, in particular, may have been influenced by social desirability bias. Though we ensured a judgment free environment, we cannot fully control for this~\cite{grimm2010social}.} The participants in our study represent only a small subset of matrimonial group users, as our Facebook group selection for the study was not systematic, which may lead to the exclusion of groups with different dynamics and user behaviors. Also, the recruitment process introduced certain limitations, as participants were selected only from private Facebook groups and through personal connections from Pakistani origin, which means results may not generalize to the entire population of the country and also other cultural backgrounds, introducing selection bias~\cite{collier1996insights}.

\section{Results}
\label{sec:results}
Here, we present our findings to answer our research question in four subsections: reasons for use and non-use of Facebook matrimony groups, information disclosure on matrimonial profiles, factors influencing (mis)trust in profiles on matrimonial groups, and expected support and protections from group admins and platforms.

\subsection{Reasons for use and non-use of Facebook matrimony groups}
\label{sec:use and non-use of facebook groups} 
\revision{Participants stated that Facebook's accessibility and cost-free nature are major drivers to use Facebook matrimony groups. Since Facebook is a popular social media platform in Pakistan that is used for various purposes, users are familiar with the features and UI, making it accessible to a wide audience\revision{~\cite{gillani2024usage}}. \rev{Most} participants used Facebook groups as a primary medium for looking for partners, and \rev{a few} others mentioned casually browsing these groups with no serious intent to make connections. Below, we discuss participants' in-depth reasoning and intent to use Facebook matrimony groups.}

\boldparagraph{Access to a large and diverse pool of potential matches in Facebook matrimony groups} 
Participants brought up numerous reasons for using Facebook matrimony groups, including Facebook's global reach. As P01 stated, \textquote[P01]{Facebook is one social media platform that is used by all the people around the world. So it was good for finding people from different countries}. Similarly, P23 highlighted the potential to interact with individuals from diverse demographic backgrounds:\textquote[P23]{They [match seekers] have not found anyone from their own society, so they try to talk to other multicultural people to interact and communicate with them}. 
In line with having access to more people, participants emphasized that Facebook's ``large audience'' (P04) and ``big community'' (P15) provided more options compared to geographically filtered dating apps.
While \rev{most} joined these groups to look for immediate serious relationships, \rev{a few} of the participants stated that they were just browsing more out of curiosity for potential matches, or \textquote[P02]{to know what kind of partner people are wanting}. Similarly, P01 shared that it's fun and interesting to read the profiles of other people who are trying to find a match and the comments. Our findings suggest that participants appreciated Facebook matrimony groups for their broad reach and diverse user base to find potential matches.

\boldparagraph{Affordability and accessibility of matrimonial groups drives usage}

The cost-saving aspect of Facebook matrimony groups is one of the benefits, as P12 shared a personal experience where a \textit{Rishta aunty}\footnote{Rishta aunty is a term used to describe women who propose potential marriage partners in return for a fee.} charged a family member around 35,000 Pakistani Rupees\footnote{At time of interview, 30,000 Pakistani Rupees is approximately equivalent to 107.94 USD, which is the minimum wage in Pakistan per month.}, which she found shocking; hence, she resorted to using Facebook. 
Participants P06, P08, and P13 highlighted both cost and ease of access with Facebook groups: \textquote[P08]{Online platform is free of cost. They do not charge you anything \textelp{} you can directly reach to the people whom you are interested}. Participant P19 specifically addressed the restrictions on other platforms, where there are restrictions on the number of people one can contact: \textquote[P19]{They reserve their count number on how many people they contact so that they won't be out of reply options, so that's why there people are reserved. On Facebook, if there are 100 members, everybody has the option to reply to anyone because Facebook is the only application that is not paid. That is why we get more responses here}. Participant P22 leaned toward Facebook’s accessibility when searching for a suitable match: \textquote[P22]{It is easy and convenient; it is like online shopping. You can easily see anyone, and anyone can approach you}. Our findings suggest that paid matrimonial platforms and apps seem less appealing because of the smaller number of users and the fact that one has to pay to access a broader pool of potential partners. \revision{This concern is reflected in media reports highlighting the need to pay for features such as profile visibility and messaging on dating apps~\cite{banfield2024}.} 

\boldparagraph{Facebook matrimony groups offer inclusive spaces}
Participants mentioned that it is not always possible to find a partner through traditional methods due to certain prevailing stigma and discrimination based on factors like colorism, casteism, ageism, and other socioeconomic factors; therefore, they prefer Facebook due to access to more matches. For example, P05 shared that \textquote[P05]{it's not always possible to find someone around [offline]}. Similarly, P07 shared their friend's experience of discrimination, \textquote[P07]{Their family had been searching for a proposal [a potential match] for two to three years without success, which led her to join these groups}. Participant P13 valued the freedom in terms of diversity of people and less stigma in Facebook groups, and preferred them over offline processes: 
\textquote[P13]{It just feels like it was an easier option, and a free option, I would say, compared to going through Rishta aunties, having a bit more freedom in looking at what you want. Other than that, like, I don't know\textelp{} just felt like it was more comfortable.} Participant P21, in particular, noted that Facebook groups can provide inclusive spaces for individuals who are marginalized in the society---including women above 30, single mothers, and divorced individuals. Overall, Facebook offered participants more diversity and they prefer using Facebook due to less stigmatization.

\revision{
\boldparagraph{Privacy concerns drove \rev{a few} participants toward paid matrimony groups}
\rev{Some} participants mentioned a preference for paid matrimonial groups, associating them with better privacy, reduced risk of misuse of information, \revision{and more serious intentions}. For example, P08 believed that paid groups intend to have members genuinely seeking matches: \textquote[P08]{If they would ask for money, then I think it's a good thing; it's not a bad thing because for free, everyone wants to pass their time. But when the money comes in between anything, people usually back out}. Participant P11 also shared a similar view, underlining how premium groups are perceived to be safe because they are limited to people who really look for serious relationships and, in turn, reduce unwanted interactions with non-serious matches: \textquote[P11]{Those [paid] groups are almost safe because they pay money, and people who pay money are serious \textelp{}}. Participant P17 expressed their fear of information misuse on free groups and explained that in paid groups \textquote[P17]{data will be securely handled. So that's why I would recommend everyone to prefer paid option rather than free versions, or free groups, free Facebook groups, or free WhatsApp groups}.} Our findings suggest that \rev{some} participants view paid matrimonial groups as offering greater security, privacy, and seriousness compared to free groups.

\begin{summaryBox}{Summary: Use and non-use of Facebook matrimony groups}
    \begin{itemize}
        \item Facebook groups offer access beyond geographic or traditional limitations. Facebook groups are seen as more inclusive, offering space for marginalized individuals.
        \item Participants in cost-free groups voiced concerns about data misuse and unwanted interaction.
    \end{itemize}
\end{summaryBox}

\subsection{Information disclosure on matrimonial profiles}
\label{subsec:info shared}

\revision{In this section, we present participants' views on sharing information, the types of information they share, and their concerns with disclosing personal details. Since it is required of match-seekers to share information (see Figure~\ref{fig:man}), even if they may not want to, we discuss protective measures participants adopt to share information safely in matrimonial groups.}

\subsubsection{Cultural norms around personal information sharing in matrimony profiles}
\boldparagraph{Participants expect profiles to contain detailed information including personal details and preferences in partners}

\revision{Participants generally perceived detailed profiles as more authentic and appealing. For instance, P01 and P02 noted that profiles with detailed descriptions and pictures, along with what they seek in a partner, grabbed attention and likes on the post. Participant P01 remarked, \textquote[P01]{The two people that I contacted for myself \textelp{} they did actually post their own picture, and their profile had a very long description of what they wanted, what kind of personality they have, and what they were looking in their partner}. They further highlighted that the number of likes on a post also influences the trustworthiness of the profile~\cite{phua2016explicating}.}

Participant P13, however, said they read the profile thoroughly; they stated that their first instinct is just to read the profile. P09 added that providing contact information shows seriousness: \textquote[P09]{if they provide any follow-up information, definitely an email address that they're open to being contacted}. They also believed that anyone who joined the group had good intentions and stated that they \textquote[P09]{think there is a built-in kind of trust \textelp{} why would somebody want to sign up on a page, provide information and pictures, and then come across as being untrustworthy?}. Participant P04 highlighted the importance of profile being honest and emphasized that profiles must be \textquote[P04]{a bit more detailed, they should show a bit of emotional intelligence in their bio}. 
Our findings indicate that \rev{most} participants expect detailed descriptions, personal photos, follow-up contact information, and visible engagement on posts as contributing to the perception of authenticity and seriousness in the search of a partner.

\boldparagraph{Participants' information sharing practices and standards vary based on individual preferences}
Participants mentioned they prefer to include only \textit{basic information} in their profiles; however, the definitions of basic information often varied across participants. \rev{Most} of the participants expressed feeling comfortable sharing some or all of the following personal and demographic information: name, age, education, gender, height, profession, marital status, caste, religion, parents' occupations, number of siblings, city, interests, and their requirements for an ideal partner. Participant P04 described that \textquote[P04]{every Pakistani profile has gender, age, what kind of work we do, height, even like what our father does, what our mother does, how many siblings we have, where we come from in Pakistan, where we live now, what are our aspirations, and what we're looking for in a partner}. However, not all participants agreed on what constituted \textit{basic information}. While some were comfortable sharing details such as their name, date of birth, and pictures, others were hesitant about disclosing ethnicity, pictures, complexion, salary, email, and phone number, citing privacy reasons and potential discrimination, which we discuss in Section~\ref{subsec:concerns-info-sharing}.

\boldparagraph{\revision{Cultural norms and family involvement drive information sharing in matrimonial profiles}}

Participants' decisions on what to share in matrimonial groups are sometimes influenced by cultural, gender, familial, and religious factors---often limiting their autonomy in the information sharing process. \rev{Most} participants did not have straightforward answers regarding why they share information on matrimonial profiles, except to say that it is the norm. \rev{A few} participants believed that sharing information can help with assessing compatibility: \textquote[P16]{There is no harm in sharing the information about \textelp{} age, height, profession, and family background}. Participant P18 mentioned that being upfront about expectations from a partner helps to avoid misunderstandings later. \rev{A few} participants also noted that mentioning details like religion, caste, and language---while sensitive and potentially discriminatory---are important considerations in matrimonial profiles for them. \revision{For \rev{some} participants, information sharing decisions are influenced by their families. Participant P01 described needing permission from their elders to post their information: \textquote[P01]{I did not really have the courage to openly post because I did not have that kind of permission from the elders, from my mother to post [on matrimonial groups]}. Similarly, P03 noted the importance of considering parental input for deciding what information to share: \textquote[P03]{If our parents say that we should not include this extra information \textelp{} I will listen to my parents}.}

\subsubsection{Concerns related to information sharing and protective behaviors} \label{subsec:concerns-info-sharing}

\boldparagraph{Participants selectively share pictures due to concerns of misuse}

Participants were cautious when sharing pictures, often opting for private messaging instead of public posts due to concerns of misuse. For example, participant P13 stated that they \textquote[P13]{don't want to show [their] picture to every single person in the world}.

\revision{\textbf{Concerns:} \rev{Most} participants expressed a range of concerns related to posting pictures online, particularly in the context of privacy and social expectations.} For example, P01 and P03 acknowledged the societal and familial expectations to share pictures; they noted that, while they were not comfortable sharing pictures, they felt ``pressured'' (P03) to do so, as sharing pictures is often treated as a requirement in Pakistan. Women, in particular, face pressure to meet beauty standards and fear judgment from strangers on posts containing their pictures~\cite{malik2023marriage}. Participant P09 further explained the need for discretion, stating fears of misuse and \textquote[P09]{fear for their privacy, like their picture being used somewhere else in a compromising way}. 

\revision{\textbf{Protective behaviors:}
To mitigate concerns of picture misuse, participants P03, P06, P09, and P18 adopted practices such as avoiding public posts and sharing pictures only through private messages once they establish trust with their potential matches. As P18 stated,  \textquote[P18]{I do not add pictures, \textelp{} I have only provided to those who respond under my post}. These strategies were motivated by fears of unauthorized use and judgment from strangers, with women being particularly cautious due to societal pressure and cultural expectations.}  

\boldparagraph{Participants avoid revealing financial information due to risk of financial exploitation}

\revision{Participants avoid sharing salary, income, wealth, or asset-related details on matrimonial profiles, fearing exploitation and the risks associated with the public accessibility of their posts and information. 

For example, P02 mentioned they would never reveal their earnings on a public forum. In some cases, women withheld financial information due to cultural norms surrounding gender roles related to employment of women: \textquote[P04]{Culturally, women don't really put their income on the [matrimonial] profiles.}.  }

\textbf{Concerns:} \rev{Some} participants expressed worries about sharing financial information for privacy reasons. For instance, P15 expressed not understanding the purpose of sharing salary information, stating that it is extremely private: \textquote[P15]{Salary is a very confidential thing, something that you cannot share with everyone publicly}. Participant P02 mentioned fears of being kidnapped, blackmailed, or exploited: \textquote[P02]{It's [sharing financial information] a risk. So I think it should not be mentioned publicly where they live or how much they are earning}.

\revision{\textbf{Protective behaviors:} \rev{To address threats associated with financial disclosures, participants adopted practices similar to how they handled sharing pictures. They revealed information in personal settings and withheld financial details unless they trusted the individuals involved.}} Participant P20 stated, \textquote[P20]{I think I won't disclose salary \textelp{} obviously I would share it on a personal level, and I won't include it in a public post.} 
\rev{Participants P07 and P11 also highlighted that they avoided details about assets to avoid unwanted attention from adversaries: \textquote[P07]{I have not shared the agricultural lands that we own, I haven’t shared my number, and I contacted them on DM}.} 

\boldparagraph{Participants express concerns about fake profiles and misuse of information, including unauthorized sharing and identity theft}

\rev{Almost all} participants stated that they deliberately avoided sharing sensitive and identifiable personal details, such as their names, addresses, and photos, especially in public matrimonial groups, to minimize the risks of exploitation. Two Participants who had children, mentioned they would be apprehensive about sharing  details of their kids, such as names and pictures, that could attract adversaries and lead to misuse. They further cautioned other users not to post this information publicly. 

\textbf{Concerns:} Participants raised significant concerns related to fake profiles and potential misuse of personal information, as well as cybercrime risks involving financial aspects like money and bank details. Further, participants feared some individuals, whom P01 referred to as ``liars'', were present in matrimonial groups to deceive legitimate users for their personal amusement. Participant P05 recounted their friend's experience with a fraudulent profile that initially seemed genuine; on further inspection, they found that the person was already married and was only looking to relocate to another country: \textquote[P05]{eventually, she got to know that the guy was already married. And he was looking for an opportunity to come to Europe}. Similarly, participant P21 also recalled a deceptive matrimonial profile: \textquote[P21]{I know a man who is already married and has a child. I have seen his matrimonial profile in one of the groups \textelp{} and I felt that this is a scam}. Additionally, P18 expressed concerns about \textquote[P18]{cybercriminals who might use your pictures or your address or your real name, and they can even make a fake profile}. Participants P03 and P06 pointed out the issue of catfishing, where individuals use someone else’s photos to gain attention from potential victims.

\revision{\textbf{Protective behaviors:} 
To avoid risks of deception and identity misuse, \rev{most} participants adopted safety measures, such as withholding sensitive details in public groups, and verifying profiles of posters by checking their photos, public posts, and assessing how old the profile is to assess its authenticity}. \rev{For instance, P01 emphasized the importance of checking the pictures and public posts on their Facebook profiles} \textquote[P01]{to determine if they are a real person or someone with multiple fake identities}. Participant P02 added that newly created profiles raise suspicion: \textquote[P02]{If you see that the profile is quite old and not a new one \textelp{} then you can get an idea that this person cannot be fake}. \revision{Beyond individual screening, some participants mentioned seeking help from Facebook groups themselves to verify profiles and avoid deception: \textquote[P08]{People use [the Facebook group] to give a profile check of a person: that my sister is going to be married, and this is the person whom I want to check if he's married or not \textelp{} so people who know them, maybe their colleague, a friend, university friend, whatever, use to give a comment; so it can save a lot of lives}.}

\revision{
\boldparagraph{Participants fear that sharing personal contact and identifiable details increases the risk of harassment and abuse}
\rev{Almost all} participants were hesitant to share personal details in matrimonial groups due to privacy concerns and the potential for unwanted attention. Although \rev{a few} mentioned sharing their phone numbers and details to allow easier communication with potential matches, others remained cautious. For instance, participant P01, who posted for a friend mentioned that they \textquote[P01]{were careful not to post her [their friend's] original name, father’s name, or family name \textelp{} anything that could give away her real identity}. Participant P02 was also against sharing specific location details and suggested that \textquote[P02]{they [match seekers] should exclude their exact location. They could say they are living in this city}.

\textbf{Concerns:} Participants expressed hesitation around disclosing sensitive details like phone numbers, citing fears of personal safety breaches and misuse of personal information. Participant P11 recounted serious incidents of phone number misuse and deception, where impersonators \textquote[P11]{talk in girl’s voice and call them, and they kidnap them and ask for ransom. It has been happening since long time ago, like 5, 6 years}. Participant P01 also shared a personal incident where the conversation turned inappropriate, stating that the abuser \textquote[P01]{started talking indecent[ly] to me, like he wanted more from me, \textelp{} like nude pictures, and he wanted me to have phone sex with him}. Participants P09, P22, and P23 worried that sharing contact details increases the risk of harassment or blackmail; participant P22 was concerned that \textquote[P22]{someone can stalk you [in person]}. \rev{A few} participants further emphasized that adversaries can retrieve contact details from various sources, such as online CVs; participant P15 supported these concerns by sharing instances of contact number misuse from CVs: \textquote[P15]{We share our CVs, we share mobile numbers [in CVs], and numbers got leaked from our CVs}. 

\textbf{Protective behaviors:} To protect themselves, participants did not share full names, details of family members, phone numbers, and information that could reveal their exact identity or location.} For instance, P05 mentioned, \textquote[P05]{They [matrimony group users] should not post their names and their parents' name \textelp{} any information that clearly indicates who this person is}. Further, \rev{a few} participants preferred using Facebook as a primary medium to contact potential matches, as profiles on social media can be blocked if suspicion arises (P19). Participant P20 suggested using email or social media instead of a mobile number for contact details: \textquote[P20]{If your number is publicly shared, you get a lot of advertisement messages, and you don't know who will reach out, or who knows, maybe scammers call you. That would be an unnecessary hassle}.

\boldparagraph{Participants' fears of gendered discrimination and societal stigma discourage them from posting on matrimony groups}
 
\rev{Almost all} participants shared essential details such as their gender and marital status, as they consider this information necessary for finding a suitable match. However, sharing such details---especially for women and those remarrying---led to uncomfortable comments, including outright discrimination and harassment (P06, P14). Participants P12 and P04 expressed discomfort with sharing physical aspects on their profiles; participant P04 explained, \textquote[P04]{Complexion, weight, or income: these things can be very judgmental, and I don't feel like it fully describes a person as a whole}. She also mentioned that her parents advised her not to share her nationality to avoid attracting people who are more interested in moving abroad than in the relationship or marriage, stating \textquote[P04]{that [sharing current nationality] can attract wrong kinds of people}.

\textbf{Concerns:} \rev{Some} participants expressed discomfort with the idea that friends or relatives might view and recognize their profiles. For example, P09 explained that some individuals may avoid posting pictures \textquote[P09]{either because they're kind of conservative, traditional or because they don't want to be judged or seen}, further noting that any mishap can be traumatic in a culture of virality on social media. Participant P08 emphasized feeling uneasy about being judged on their physical appearance, particularly when others make body shaming comments on pictures: \textquote[P08]{It can either boost your confidence or lower your self-esteem}. \rev{Some} participants also highlighted cyberbullying and demoralizing comments as a major issue. For example, P06 recounted that a man was repeatedly targeted in the group for seeking a second marriage;  P03 who observed mean comments under others’ posts noted that such negativity could emotionally hurt and impact one's mental health. Additionally, P14 highlighted that young women---divorced or widowed---who have children face heightened discrimination, including sexist or inappropriate comments and real-life bullying from coworkers who view their posted profiles. 

\revision{\textbf{Protective behaviors:} 
In response to fears of judgment and social scrutiny, some participants refrained from posting in matrimonial groups entirely, some posted anonymously, or relied on family and friends to post on their behalf. For example, P01 recounted posting for their friend \textquote[P01]{because they [their friend] did not want to give their own identity, their own ID on the group, so I just helped them to post their profile}. \rev{Additionally, others chose to avoid sharing photos, physical attributes, or personal identifiers in their matrimonial posts to minimize exposure and avoid social judgment.}}

\boldparagraph{Participants verify profiles by meeting people in-person to avoid scammers}
\revision{Participants described exercising caution after identifying a potential match in matrimonial groups, often withholding personal details until they verify the person's identity. For instance, P23 emphasized that it is easy to fake identities online, highlighting the need for increased scrutiny and verification of profiles to build trust.}  

\textbf{Concerns:} \rev{Most} participants expressed skepticism toward the authenticity of online profiles, fearing scammers impersonating others online. For example, P19 recounted a personal experience where they were misled by someone whose intentions did not match what they stated on their online profile. Participants P06, P19, and P23 indicated that face-to-face meetings are essential for confirming someone's identity.

\textbf{Protective behaviors:} To safeguard against scams and misrepresentations, \rev{most} participants engaged in prolonged conversations to assess authenticity, withheld personal information until trust was established, and preferred in-person meetings---often in the presence of family members---to confirm identity before moving forward in the matchmaking process.

\begin{summaryBox}{Summary: Information disclosure on matrimonial profiles}
    \begin{itemize}
        \item Participants do not agree on what information they must share on a matrimonial profile. They prefer more information from potential matches but prefer sharing minimally on their profiles. 
        \item Participants felt compelled by cultural norms to share personal information, yet also took protective measures (e.g., limiting disclosures, avoiding picture sharing, meeting in person) to protect themselves against harassment, financial exploitation, and \revision{romance scams}\footnote{\revision{Romance scams involve cybercriminals engineering a romantic relationship on online dating platforms for monetary gain~\cite{bilz2023tainted}.}}. 
 
    \end{itemize}
\end{summaryBox}

\subsection{Factors influencing (mis)trust in profiles on matrimonial groups}
\label{subsec:factors (mis)trust}
Below, we present several insights from participants regarding the factors that shape their trust and mistrust when interacting with profiles in matrimonial groups.
\subsubsection{Factors that influence trust}
\label{subsubsec:trust factors}
\revision{\boldparagraph{Participants trust profiles with family involvement}

Participants mentioned familial influences in the entire matrimony process as well as in information sharing decisions. \rev{A few} participants mentioned depending on their families to post on their behalf (P04, P03, P01), indicating high trust in their family's judgment. Participants also indicated preferences for Facebook over other platforms like SingleMuslim~\cite{singlemuslim2023} and Muzz~\cite{muzz2015}; they noted that Facebook allows families to post on their behalf: \textquote[P04]{It's usually somebody looking on behalf of their family members, and then there is a bit more trust there}. One participant, in particular, stated that they place complete trust in profiles posted by family members: \textquote[P08]{If someone wants their family involvement, that means that the person is 100\% reliable}. Participants P10, P13, and P18 viewed early family involvement as essential for trust, with P13 calling it an added layer of security. Participant P21 also noted, \textquote[P21]{When people involve their family in the initial stage \textelp{} you can get to know that person and their family better, and there are fewer chances of wasting time or encountering scams}. Along the same lines, P18 highlighted early family involvement as a sign of commitment: \textquote[P18]{The involvement of the family is very crucial from the very initial state. This shows the seriousness of the opposite party}. 

\rev{Some} participants described how specific posts by or about family boosts their confidence in a profile: \textquote[P04]{It [family involvement] creates a bit of trust that it's just not some random guy on the dating apps, like Salams~\cite{salamsapp} or Muzz~\cite{muzz2015}, who is not serious \textelp{} if a family's contacting another family, there's more trust}. Participant P05 stated that \textquote[P05]{if somebody has mentioned clearly that my family will also be involved \textelp{} I can easily contact that person because I would say that this person is serious. But if somebody has not posted anything like that, it doesn't mean that the person is not serious, but I will surely have a doubt}. P14 and P19 shared their gradual approach to involving family; P19 noted that \textquote[P19]{the first thing is that the boy or girl talks to each other, and then families should be involved}. 
Therefore, we find that family involvement in matrimonial profiles signals trust and indicates seriousness in the search for a partner}.

\boldparagraph{Participants use cross-platform verification of profiles (like LinkedIn, Instagram) to establish trust} 

\rev{A few} participants mentioned cross platform verification across other social media platforms as a strategy to assess the authenticity of a profile. For example, P03 shared their process of verifying a profile: \textquote[P03]{We can add them on Snapchat and ask for their live photos, or we can also go to their followers there, we can check their friends \textelp{} and ask about the person}. They further added that they would definitely use platforms such as Instagram to verify whether the person is real.
Participants P09 and P17 mentioned how they use LinkedIn to verify the authenticity of profiles and one \textquote[P17]{can also get all the information, like from where the person has done schooling, college, and the graduation. If person is currently working, then which company he or she is currently working at. Then you can also find the company members, you can also send a message to them and ask the person}. Our findings indicate that participants prefer to verify profiles further through interactions on other social media platforms and by reviewing professional and other personal information.

\subsubsection{Factors influencing mistrust}

\label{subsec:mistrust factors}
\boldparagraph{Participants raise concerns about the authenticity of
anonymous posts} 

The anonymous feature on Facebook~\cite{anonymous-facebook} allows members to post in groups without revealing their identity, using an anonymous label or a nickname; this feature signaled mistrust in some participants. Only group admins and Facebook administration can see who they are---not other members---as shown in Figure~\ref{fig:anonymous} in the Appendix. Prior research highlights the importance of anonymous communication in protecting users from privacy violations and identity theft~\cite{hoang2014anonymous}, and \rev{some} participants in our study echoed these views. They had mixed reactions toward anonymous posts---while some valued the privacy they offer, others expressed mistrust: \textquote[P11]{If someone doesn’t want to show their identity, then there must be a reason \textelp{} I won’t trust them}. Few participants also questioned the reliability of anonymous posts, especially when there is no way to confirm who is responding: \textquote[P14]{People post their profile anonymously, and they don't share their numbers nor links to their accounts. When you post a comment under their post, then someone sends you an inbox [message], but how would you know that it's the same person that you commented under their post?}.

\rev{A few} participants shared how they navigate anonymous interactions. For instance, P17 explained their process for engaging with such anonymous posts, stating, \textquote[P17]{we need to send them [potential matches] a message, we need to get the information, get the contact number, then we need to meet. Personally, then we'll find out, if the person is real or fake}. Participant P20 offered a more flexible view on trusting anonymous posts and \textquote[P20]{as long as the post is detailed and every information is included and is interesting then I would like to respond to that post; it does not matter if the post is anonymous}. 
In contrast, \rev{some} participants appreciated the option of posting anonymously. Participant P08 mentioned that people choose anonymous posts for mental sanity: \textquote[P08]{I think the anonymous option is a great option. And it should [remain] forever. So nobody knows that who is posting}. Similarly, participant P13 stated, \textquote[P13]{There is a lot of shame associated with looking for a spouse, especially for girls \textelp{} so having anonymous option is a really good thing for them}. This shame stems from societal stigma toward women seeking a partner on their own~\cite{akram2024stereotypical}. Our findings indicate that while \rev{some} participants value the anonymous posting feature, others express concerns about the authenticity of anonymous profiles, raising the need for better anonymous post verification mechanisms.

\boldparagraph{Participants raise concerns about the authenticity of newly created, lavish, and restricted profiles}

\rev{A few} participants expressed concerns about the authenticity of newly created profiles: \textquote[P03]{if someone has created a new profile just to propose [marriage] \textelp{} then it looks like it's [a] scam or that he or she wanted to just \revision{make friendship\footnote{\revision{a term used in Pakistan to indicate establishing a relationship without the intent of marriage~\cite{mahboob2008pakistani}.}} or for timepass\footnote{\revision{a common term used in South Asia, especially Pakistan, to indicate passing time aimlessly~\cite{parker1998culture}.}}}}. In addition to concerns about new accounts, P03 and P08 highlighted that Facebook’s profile lock feature makes it harder to assess the credibility of profiles, therefore, \textquote[P03]{\textelp{} it's very difficult now to trust people on Facebook because we can't find out if the ID is new or not}. Participants also viewed some profiles with suspicion due to their exaggerated wealth and financial information. 
Participant P07 echoed this sentiment, extending concerns to profiles that displayed excessive wealth or material possessions: \textquote[P07]{Some people show off [wealth and assets] a lot. If they have a car, they also mention it in their profile that they have a car. So, from these things, you can find out that they show off a lot}. Participants indicated that their suspicion extended to the amount of information in a profile, noting that both too little and too much information could be an indicator of inauthenticity: \textquote[P03]{With incomplete details \textelp{} maybe people are not interested, but the person will not get many responses because of missing details. On the other hand, if you include irrelevant or [too] much information in the bio-data, then it will be misused by people}. They further added that some users seemed more interested in casual friendships than in pursuing serious relationships. 
Therefore, participants remain cautious of profiles with limited or locked information, new accounts, and those that showcase excessive material wealth, as these factors often raise suspicions about the user’s intentions and authenticity. 

\boldparagraph{Participants signal mistrust toward matchmakers who monetize their services}
\revision{\rev{A few} participants also expressed concerns about admins of Facebook groups who sometimes monetize their role and act as fraudulent matchmakers. Participant P14 mentioned that some admins turned groups into businesses by charging fees and collecting identification documents for matchmaking purposes, indicating that these practices may be scams and are untrustworthy.} Participant P04, who is an admin herself, stated that she removes fake matchmakers from her matrimonial groups \textquote[P04]{because most of them [matchmakers] are scammers, they take your information, there's no liability, there's no reputation for them. They can just, like, disappear tomorrow, make another account, and post under a different company name}. \revision{She further explained how these fake matchmakers often post profiles that appear too good to be true, demand payment from members to connect with them, and then disappear once they receive money.}

\begin{summaryBox}{Summary: Factors influencing (mis)trust in profiles}
    \begin{itemize}
        \item Profiles with detailed personal or family information (including those posted by family members) tend to reinforce trust, while anonymous or incomplete profiles often raise suspicions of romance scams. 
        \item (Fake) Matchmakers who monetize their services and demand documents from users raise concerns. 
    \end{itemize}
\end{summaryBox}

\subsection{Expected support and protections from group admins and platforms}
\label{subsec:role of admins and platforms}
Below, we discuss participants’ expectations regarding the support and safety measures they believed group admins and the Facebook platform should provide in matrimonial groups. We also discuss participants' dissatisfaction with the current role of administrators in Facebook matrimony groups, highlighting the need for stricter policies from admins and the platform.
\subsubsection{Expected role of matrimonial group admins} 
\label{subsubsec:role of admins}

\boldparagraph{Participants suggest admins should enforce stricter content moderation policies for fake profiles, scammers, and cyberbullies}

\rev{Most} participants recommended that admins take a more active role in removing unserious members, lurkers, and abusers from the groups. For example, P01 shared a personal experience, where an admin failed to take action against someone they reported: \textquote[P01]{I texted the admin that this guy is actually not serious and he talks really inappropriately. So they asked me for proof, but by that time I had already deleted all the chats that I had with that guy. So I literally had no proof}. She further added that admins typically do not interfere unless someone complains or reports. Similarly, P22 believed that admins should take action when a group member raises a concern. Participants P08 and P18 supported stronger admin involvement for safety and verification: \textquote[P08]{If the admin steps in to verify whether a person is genuine, then there would never be any problem}. Participant P21 shared that they \textquote[P21]{have mostly seen that admins don't take action, [the] maximum that they do is delete the post and also that happens rarely}. This indicates that there is a mismatch in users' expectations from matrimony group admins and what admins actually do.

\boldparagraph{Participants prefer admins to provide security and privacy suggestions in groups}
\rev{About half} of the participants emphasized the role of admins in promoting security and privacy within matrimonial groups. For example, P20 and P16 recommended that admins create detailed, pinned posts for new members to review before participating: \textquote[P20]{\textelp{} so that anyone who wants to join the group can see it. When someone wants to join, they first fill the questionnaire and read the group rules}. Similarly, P04 suggested that admins post recurring reminders to encourage safe behavior, and suggested that \textquote[P04]{they [admins] can put a weekly post [for group members] \textelp{} on how to protect themselves and to report if they see anything suspicious}. Participants also appreciated when admins cautioned users against oversharing with others, \textquote[P05]{unless you know a person a little bit more}. Participants P03 and P06 further recommended that admins provide a proper template for posts, as appended in Figure~\ref{fig:template}, which they believed would help keep the group secure by ensuring posts follow clear and consistent rules. Additionally, P08 stressed the need for admins to raise awareness about scams and their legal consequences: \textquote[P08]{They should have to give the awareness, post something, and motivate them that if there's something happens, if there's any scammer, we should report it to the police}.

\boldparagraph{Admins lack the capacity to verify authenticity of posts and moderate content}

Participants expressed clear expectations about the responsibilities of admins in matrimonial groups but also acknowledged the limitations admins face in the verification process. For example, P12 mentioned, \textquote[P12]{It's very hard to find out if someone is fake or not because there are thousands or you know can say millions of fake IDs [Facebook profiles] doing some sort of negative stuff on the Facebook}. Similarly, P04 emphasized the practical constraints involved:  \textquote[P04]{They're not making any profit. Imagine that these groups can have 100,000 members. It's simply not viable to expect one person or a few people to vet every single person}. Participant P13, another admin, echoed these challenges and explained that although admins try to screen members, they face clear limitations in verifying authenticity of profiles and cannot fully moderate content or membership. 

\rev{A few} participants were critical of admins who prioritized growth over authenticity and felt that admins should be accountable if data is misused. For instance, P13 remarked, \textquote[P13]{Just to increase the numbers [group size] and have big numbers doesn't mean that they are more authentic. This just means that they're adding a lot of people without even verifying them at all}. Similarly, P19 shared a more skeptical view, highlighting the limited impact and involvement of admins: \textquote[P19]{The admins' role is not so significant, nor do they have any interest, because thousands of posts are made in the group, and there are a limited number like approximately 3 [to] 4 admins}. Our findings suggest that while participants also recognize the challenges admins face due to large group sizes and limited resources; this highlights the need for better safety mechanisms from the Facebook platform itself.

\subsubsection{Expected role of the Facebook platform}
\label{subsubsec:role of platforms}

\boldparagraph{Participants suggest platforms to restrict screenshot and image download options for privacy reasons}

Participants P01, P04, and P08 indicated that they would be more willing to post pictures if platforms prevented screenshots and image downloads. Participant P10 acknowledged the limitations of fully blocking screenshots, since users can still use other devices to photograph screens or browser tools to save pages, but noted that Facebook \textquote[P10]{cannot stop that [taking screenshots], but \textelp{} yes we can stop image downloading}. Similarly, P12 proposed that platforms like Facebook introduce group level settings to block screenshots, stating, \textquote[P12]{that no one can take screenshots}. Our findings show that participants strongly desire privacy focused platform features to protect their images and sensitive information.

\boldparagraph{Participants expect better profile verification mechanisms and signifiers}
\rev{Most} participants highlighted different ways platforms could verify profiles and enhance trust within groups. For example, P09 suggested implementing a verification badge system similar to Instagram's blue check mark~\cite{dumas2021importance} to indicate that a group member's identify had been authenticated: \textquote[P09]{They could have a checkmark beside them saying that this is the verified identity. Then, perhaps, people will be more comfortable}. Participant P11 also emphasized the importance of identity verification and cautioned against trusting group members too easily, as accounts can be created with little effort. They further added that dating apps provide a greater sense of security through verification mechanisms: \textquote[P11]{you get [a] verification code; only then you can create an account}. Similarly, P19 highlighted how verification through email and phone numbers on dating apps reduces the risk of scams. Participants P13 and P23 discussed the value of formal identity verification using official documents, mentioning that some groups already request items such as \textquote[P13]{your ID, your [educational] certificates \textelp{} they would ask your employment letters}. Taking it further, P23 proposed integrating document-based verification directly into Facebook's platform: \textquote[P23]{Proper verification should be done either through [an] ID card or a passport}. Our findings indicate that participants prefer a strong, multi-step verification process to increase trust and reduce scams in matrimonial groups.

\boldparagraph{Participants recommended adding search filters for factors like age or location for better matchmaking}

\rev{A few} participants expressed frustration over the lack of search and filtering options in Facebook groups, suggesting that filtering features such as age, location, and preferences would make the matchmaking experience more focused and efficient. For instance, P04, who had experience posting both for herself and on behalf of her parents, explained, \textquote[P04]{There aren't enough features in Facebook groups to do all the filtering and the searching. It's just whatever comes into your feed}. She emphasized that, in contrast, dating apps offer more control and a structured experience through filtering systems, adding, \textquote[P04]{I do wonder how people find the right people when there isn't a filter that you can search through}. Similarly, P14 recommended adding filters based on age to improve the search process: \textquote[P04]{There should be a filter so people can search profiles on age-based}. 
\begin{summaryBox}{Summary: Expected support from group admins and platforms}
    \begin{itemize}
        \item Participants expected admins to remove harmful comments, proactively verify user legitimacy, and provide safety guidelines and scam awareness.
        \item Participants suggested implementing verification systems (Phone, email, ID) to reduce the risk of fraud. 
    \end{itemize}
\end{summaryBox}
\section{\rev{Discussion and Recommendations}}
\label{sec:discussion}
Through 23 semi-structured interviews, our findings reveal the complex interplay between cultural norms, privacy concerns, and the use of social media among Pakistani users. In this section, we discuss our findings and present recommendations for each stakeholder in the matchmaking process.

\subsection{\rev{Discussion}}

\subsubsection{Cultural and social factors influencing trust and information sharing}
\boldparagraph{Self-disclosure asymmetry in matrimony profiles: mismatch in information disclosure practices vs. expectations}
Users in online communities exhibit a \textit{self-disclosure asymmetry}---they are hesitant to share personal details but tend to trust profiles that include more personal information~\cite{pechmann2017self}. Our study findings reflect similar trends, as presented in Sections~\ref{subsec:info shared} and ~\ref{subsubsec:trust factors}. Participants expressed reluctance to disclose personal information due to concerns about misuse, identity theft, and social judgment; however, they were more likely to engage with profiles that included detailed information, viewing these as belonging to individuals with serious matchmaking intentions. 
\revision{This behavior reflects the global `privacy paradox'---a difficult trade-off in which sharing more personal data may improve a profile's appeal and increase responses but also exposes users to greater risks~\cite{taddicken2014privacy}.}
Additionally, prior research shows that personal data---especially in online dating contexts---can be weaponized or exploited if collected and used without proper safeguards~\cite{gulyamov2023personal,warner2018privacy}. To strike a balance, participants recommended proceeding with caution by reaching out directly to potential matches and sharing personal details privately to assess trustworthiness, rather than posting them publicly (Section~\ref{subsec:info shared}).

\revision{\boldparagraph{Influence of family involvement in trust building aspects in Pakistani society} Participants in our study emphasized that family involvement in the matchmaking process is highly valued and enhances trust. Prior research similarly notes that, although arranged marriages have declined, they remain common in Pakistani communities~\cite{menon1989arranged}. Participants noted that posts involving family members were perceived as more credible and less likely to be scammers (Section~\ref{subsec:factors (mis)trust}), aligning with prior work showing that in family-oriented cultures, parents actively guide and validate their children's partner choices~\cite{buunk2010cultural}. Such influence remains significant in South Asian communities, where families may not always directly arrange marriages but still shape partner selection by influencing social interactions and expectations~\cite{talbani2000adolescent}. While \rev{some} participants valued and respected family input, others asserted their own preferences and autonomy in information sharing practices---highlighting a balance between individual choice and family influence}. However, we emphasize that higher trust in profiles posted by family members can create more opportunities for scammers to pose as trusted family members and lure unsuspecting victims.

\subsubsection{Security and privacy risks associated with using Facebook matrimony groups and online dating platforms}

\boldparagraph{Anonymous posting and its dilemma in matrimony profiles}
While the anonymous posting feature is designed to protect user privacy in Facebook matrimony groups, it often creates a dilemma for both users and viewers in terms of whether the post is legitimate~\cite{wang2017onymity}. This feature allows individuals to navigate social stigma and avoid judgment without revealing the posters' identity; however, it also raises concerns about authenticity, as participants reported that the lack of identifiable information made it difficult to assess the users' intentions. They also expressed concerns where scammers can fabricate random details that may look attractive to others and misuse this feature to lure potential victims (Section~\ref{subsec:factors (mis)trust}). \revision{Additionally, our findings suggest that anonymous posts do not fully conceal the identity of legitimate posters, as matrimony posts often include identifying information that can be traced back to an individual (Section~\ref{subsec:info shared}). Although the anonymous posting feature may offer the benefit of concealing one's identity, our findings---consistent with prior research---show that most participants were reluctant to engage with such posts due to concerns about fake profiles, misuse of shared information, and lack of accountability. This suggests that, the risks outweigh the benefits and highlights the need for better mechanisms to distinguish legitimate users from scammers~\cite{binns2013facebook}.}
Participants who do not use the anonymous posting feature progressively disclose personal information, rather than all at once, a strategy shown in prior work to reduce uncertainty and build trust~\cite{knobloch2008uncertainty}.

\boldparagraph{Risk of personal information misuse, identity thefts, fake matchmakers, and scams}
Participants cited data security and privacy as major concerns for using Facebook groups, particularly regarding the misuse of personal information and pictures (Section~\ref{subsec:info shared}). \revision{These concerns reflect similar risks identified in prior work, such as the misuse of personal data leading to harassment and fraud~\cite{odusote2021data,kroger2021data}. In our study, \rev{some} participants, including group admins, raised concerns about fake matchmakers impersonating others to scam group members. This concern emphasizes the need for stronger admin controls, ID-based account verification, trust-badges, and better mechanisms to report scammers (Section~\ref{subsec:role of admins and platforms}).} Participants also reported instances of catfishing, where deceptive profiles used stolen photos or fabricated details to appear more trustworthy, reflecting findings from previous studies~\cite{soneji2024feel}. 
To address deceptive behavior and enhance the protection, participants recommended technical features such as disabling screenshots and downloads within group settings to better protect personal photos (Section~\ref{subsubsec:role of platforms}). These suggestions align with prior research, which highlights how the unrestricted screenshot feature undermines users' willingness to share sensitive content~\cite{shore2023platform}. Participants also supported face-blurring tools to enhance privacy, aligning with prior work that proposes automatic face blurring to control photo visibility without consent~\cite{ilia2015face}.

\boldparagraph{Inevitable exposure to privacy invasions, harassment, and physical security threats}
Participants highlighted fears of imminent threats of privacy invasions like identity thefts, doxxing, stalking, harassment, and even misuse of pictures. These fears also seemed gendered, with heightened risks faced by women, who also mentioned balancing cultural values and judgment from the society (Section~\ref{subsec:info shared}). \revision{The threats align with prior research on online harms including romance scams, identity theft, social engineering, and cyber bullying~\cite{cole2024qualitative,soneji2024feel}}. \rev{About half} the participants recounted personal stories of themselves or people they knew who were scammed or at least vulnerable to scammers in the past (Section~\ref{subsec:info shared}).
Furthermore, participants mentioned heightened risks to physical safety, including fears of kidnapping and financial extortion. These concerns reflect broader risks identified in prior research on romance scams~\cite{soneji2024feel,chaudhry2024,down2018fia, cole2024qualitative}.

\subsubsection{Security and privacy advice}
\boldparagraph{Lack of security and privacy advice tailored to Facebook matrimony group users}
Participants emphasized the need for group admins to provide safety advice, implement stronger safety measures, such as user verification, stricter moderation of content, and clearer guidelines to create more secure environments within matrimony groups (Section~\ref{subsec:role of admins and platforms}). While Facebook does offer safety advice, there is a need for such platforms to make their safety advice more accessible and tailored to diverse user groups and specific use cases~\cite{facebookPrivacySecurity}. In our study, participants also highlighted that family members frequently post matrimonial content on behalf of their children or relatives. \revision{Participants---especially women---recounted relying on older, and often male family members to post on their behalf, which reduced their autonomy over their personal information. A lack of control on one's personal information may increase susceptibility to scams and other online harms. And often intermediaries, like older family members, may struggle to assess what information is appropriate to share and may share more than what is necessary (Section~\ref{subsec:factors (mis)trust}). This reflects prior research showing that older adults, who are often not digital natives, face challenges navigating privacy settings and identifying harmful content~\cite{kemp2023consumer, bailey2021older}. Therefore, our findings highlight the need for better safety education with respect to information sharing practices.}

\subsection{\rev{Recommendations}}
Here, we present our recommendations for Facebook matrimony group users, group admins, platforms, policymakers, and future researchers to improve safety and privacy within Facebook matrimony groups. We suggest limiting access to sensitive information, introducing ID-based profile verification, enabling filtering tools for better matchmaking, restricting fake accounts, and enforcing effective content moderation practices. While the onus is often on the users and admins to safeguard themselves from harm, we urge policymakers and Facebook to enforce policies and features that better prevent misuse and abuse.
\boldparagraph{Recommendations for users}
Based on insights from the participants, we recommend several safety measures for users engaging in Facebook matrimony groups to enhance their safety and privacy. We recommend users to avoid disclosing sensitive information, such as full names, contact numbers, birth dates, and exact locations in public posts. When sharing photographs, blurring faces or avoiding sharing them altogether is advised to prevent misuse (Section~\ref{subsec:info shared}). We also recommend using privacy settings to not disclose sensitive and private information on Facebook profiles. Following admin provided templates can help ensure posts remain concise and avoid oversharing (Section~\ref{subsubsec:role of admins}). Following the principle of progressive disclosure, we encourage users to share personal information gradually, privately, and only after trust has been established~\cite{knobloch2008uncertainty}. We encourage participants to take proactive privacy precautions while interacting with strangers, as scammers may pose as legitimate users. Also, we strongly advise users to remain cautious of unregistered or fake matchmakers, who often request payments and disappear without delivering promised matches.

\boldparagraph{Recommendations for group admins}
Based on our findings in Section~\ref{subsubsec:role of admins}, we recommend that admins carefully monitor anonymous post submissions to ensure they are used responsibly and do not become a medium for harassment or deception~\cite{mclaughlin2012norm}. Participants emphasized the need for proactive moderation, including thorough investigations, removal of fake profiles, monitoring for body shaming, and flagging inappropriate comments. However, manual content moderation in large groups, with thousands of users, may be impractical. Therefore, we encourage providing templates and guidelines for posts enabling only minimal information disclosure and only accepting posts that follow their templates. We also recommend frequent monitoring for hateful comments and harassment, and reporting them to the platform or to legal authorities.

\boldparagraph{Recommendations for the Facebook platform} 
To enhance safety, trust, and usability in matrimonial groups, we suggest several improvements that the Facebook platform could consider, many of which align with concerns raised by our participants (Section~\ref{subsubsec:role of platforms}). To prevent misuse, participants expected that users who repeatedly violate group rules should be restricted from the platform (Section~\ref{subsec:role of admins and platforms}); however, adversaries can still create multiple fake accounts and continue to misuse the platform.  Therefore, a key recommendation from participants is to strengthen identity verification---ideally at the time of account creation, or at the latest while gaining access to use matrimony groups---by requiring official ID documents that match the profile name. This measure can reduce impersonation and ensure profile authenticity. \revision{Similar verification mechanisms exist in dating apps like Tinder and Bumble~\cite{lannerhjelm2024love}. Features like blue ticks or verification badges, as seen on Instagram~\cite{deursen2023effect}, could further reinforce trust. However, the blue tick feature is currently available as a paid service by Meta; scammers may misuse the badge to falsely gain trust from potential victims~\cite{BlueTicks}. 
We recommend the platform to enforce stricter moderation in the account verification process, to ensure only genuinely trustworthy profiles get a verified status.}

\boldparagraph{Recommendations for designing Facebook matrimony groups}
\revision{Beyond identity verification, participants emphasized a need for implementing stronger content control features such as disabling screenshots or downloads~\cite{cobb2017public} and offering optional face-blurring---similar to what is implemented in apps like Tribal~\cite{tribalapp2025}, to protect photos and reduce the risk of misuse.}
While Facebook currently applies image compression, this alone is insufficient for protecting sensitive content~\cite{hiney2015using}. \revision{Participants also tended to trust the profiles or posts involving family, so we recommend features like tagging family members, posting on behalf of relatives, or adding a family-supported tag to signal authenticity, reflect family support, and align cultural expectations around matchmaking~\cite{yeung2018families}. Unlike mainstream dating apps, these culturally specific trust signals are unique to Pakistani matrimony groups and are not currently addressed by existing platform features.}

\boldparagraph{Recommendations for policymakers}
In addition to platform-level improvements, participants strongly emphasized the need for government intervention to create a safer digital matchmaking environment (Section~\ref{subsubsec:role of platforms}). \revision{Current laws lack specific provisions for online matrimonial groups, leaving gaps that enable the risk of misuse~\cite{akhlaq2021cybercrime, zahid2024cybercrime}. Additionally, enforcement issues---such as limited public awareness and unclear reporting procedures---further reduce the effectiveness of existing laws and regulations, highlighting the need for targeted cybercrime laws focused on digital identity misuse and online deception to better protect users~\cite{khan2022cyber,bokhari2023quantitative}.} 
Participants suggested that treating group rules as legally enforceable could encourage more responsible behavior, reflecting broader models that highlight how penalties and incentives can motivate stronger platform-level data protections~\cite{wang2023evolutionary}. However, prior research also cautions against the risks of overregulation, which can inadvertently suppress user freedom~\cite{common2023beyond}. To balance these concerns, we recommend that cybercrimes policies be clearly defined, user-centric, and designed to protect digital rights without compromising freedom of expression.

\boldparagraph{Recommendations for the security research community}
Our findings highlight several key directions for future research aimed at improving safety, trust, and user engagement on online matrimonial platforms, particularly within Facebook groups. One key area involves examining the impact of identity verification features, such as verification badges or document-based confirmation, on users’ willingness to engage and disclose personal information. Another promising direction is the evaluation of privacy-enhancing features like face-blurring and watermarking, which can help users regain control over their visual data and reduce the risk of misuse. Additionally, future work can explore flexible anonymity options, such as enabling private comments or anonymous replies to anonymous posts, and could reveal how these features affect participation rates and trust dynamics. We also recommend exploring the policies implemented by Facebook and other matrimonial platforms, including their mitigation processes, reporting mechanisms, and user support systems, to understand how these policies help to prevent data misuse and create a safer online environment. Investigating these interventions in different cultural contexts, provides a broader perspective on the effectiveness in promoting online safety and trust.

\section{Conclusion}
\label{sec:conclusion}
In this study, we explored how Facebook matrimony group users from Pakistan navigate security and privacy concerns. Through 23 semi-structured interviews, we observed that \rev{most} participants were aware of potential risks such as identity theft, misuse of personal information, pictures, and fake profiles. However, the motivation to maintain privacy often conflicted with the need to present detailed, appealing profiles to improve matchmaking outcomes. This created a conflict between posting a profile and the fear of vulnerability to scammers, harassment, and sometimes bullying.
Although participants adopted protective measures---such as limiting shared details, cross-platform verification---their ability to protect themselves was sometimes constrained by the limitations of Facebook’s group infrastructure, including the lack of identity verification, profile filtering, and picture protection. While some admins attempted to mitigate risks by offering guidelines and removing harmful content, participants believed more proactive moderation and clearer privacy controls were needed to ensure safer interactions on the platforms. Participants also highlighted the need for technical enhancements such as restricting screenshots and picture downloads, and suggested features like anonymous but traceable communication options to facilitate a safer environment. In addition to these platform-level improvements, participants stressed that stronger cybercrime regulations and public awareness campaigns are vital to creating a more secure digital matchmaking process. 

\section*{Acknowledgments}
We want to express our deepest appreciation to Anna Lena Rotthaler and Rachel Gonzalez Rodriguez for their feedback and interest in this research. We thank our participants for providing valuable input during the interviews. We also thank all the anonymous reviewers and the shepherd for their thoughtful comments and feedback.

\bibliographystyle{ACM-Reference-Format}
\bibliography{bib}
\newpage

\appendix

\section{Participant Recruitment Survey}
\label{tab:Participant-recruitment-survey}
\footnotesize
\begin{enumerate}
    \item Have you ever posted in a matrimonial Facebook group?\\
    a) Yes \\
    b) No \\
    c) Other: [Free text field] 
    \item Did you post for yourself or someone else (e.g., a friend or family member)? \\
    a) For myself \\
    b) For someone else \\
    c) I have not posted
    \item What is your age? \\
    a) 18-24 years old \\
    b) 25-34 years old \\
    c) 35-44 years old \\
    d) 45-55 years old \\
    e) 55 years or older
    \item What is your gender? \\
    a) Man \\
    b) Woman \\
    c) Non-binary \\
    d) Prefer not to disclose \\
    e) Other:
    \item What is your education level? \\
    a) Primary Education \\
    b) Secondary Education \\
    c) Bachelors Degree \\
    d) Masters Degree \\
    e) Doctoral (Ph.D.) \\
    f) Prefer not to disclose
    \item Do you have a degree in computer science? \\
    a) Yes \\
    b) No \\
    c) Other: 
    \item What is your current employment status? \\
    a) Employed full-time \\
    b) Employed part-time \\
    c) Self-employed/Freelancer \\
    d) Not working \\
    e) Not working (retired) \\
    f) Prefer not to disclose \\
    g) Other: 
    \item What is your income level per year (in Pakistan currency PKR) ? \\
    a) Up to 2.48 lakh PKR \\
    b) Between 2.48 lakh and 8.45 lakh PKR \\
    c) Above 8.45 lakh PKR \\
    d) Not applicable \\
    e) Prefer not to disclose
    \item What is your current marital status? \\
    a) Single \\
    b) Married \\
    c) Divorced \\
    d) Widowed \\
    e) Separated \\
    f) Prefer not to disclose
    \item Where do you currently reside (country)? \\
    a) Please select from the drop down
    \item Are you willing to participate in a follow-up interview to discuss your experiences and perspectives on privacy and security concerns in matrimonial Facebook groups? \\
    a) Yes \\
    b) No
    \item If you are willing to participate in an interview, please provide your preferred contact method (e.g., email, Facebook ID, phone number) for further communication. \\
    a) Please enter the communication details
\end{enumerate}

\section{Interview Guide}
\label{tab:interview-guide}
\footnotesize
\textbf{Reasons to use matrimonial groups}
\begin{enumerate}
    \item Could you tell me whether you posted a matrimonial profile for yourself/ [ for someone else]?
    \item Can you describe the motivation behind creating a matrimonial profile on Facebook for yourself/ [ the person you posted for]?
    \item Why do you use Facebook for matrimonial purposes? (why not dating apps or other matrimonial websites)
    \item What other platforms do you use?
    \item Do you have a preference for using any of these platforms?
    \item Why do you prefer using Facebook for matrimonial purposes over other platforms?
    \item How did you first learn about matrimonial groups on Facebook, and what made you join them?
    \item What kind of information did you include in your/ [the person you posted for] matrimonial profile and why did you include this information?
    \item What kind of information did you exclude/ refrain from sharing specific information from your/ [ the person you posted for] matrimonial profile and why?
    \item How do you decide what information to share (at all) publicly vs privately?
    \item Do you know anyone personally who got married through these groups?
\end{enumerate} 
    \textbf{Reasons for sharing/ not sharing information} 
\begin{enumerate}
  \item Does your/ [ the person you posted for] family know that you are a member of a matrimonial group?
    \item What role does your/ [the person you posted for] family play in your search? How do their expectations affect what you share online?
    \item Does your/ [ the person you posted for] family influence you to share specific information? If yes, what and why?
    \item What information has your family requested you not to share online, and why?
    \item Has your/ [ the person you posted for] family shared stuff about you/ [them]?
\end{enumerate}
    \textbf{Threats associated with matrimonial groups}  
\begin{enumerate}
    \item What potential risks should one consider when disclosing personal information on Facebook?
    \item  What privacy risks do you fear when sharing personal information on Facebook?
    \item How do you deal with this when facing unpleasant emotions, like being harassed or disappointed?
    \item Can you share any experiences where you've dealt with unpleasant situations online during the matrimonial search?
    \item What motivates you to continue being a part of these groups?
    \item What are the risks for the people who post online?
\end{enumerate}
    \textbf{Trust factors} 
\begin{enumerate}
    \item How do you determine if a Facebook matrimony profile/post is genuine?
    \item Can you provide additional examples of cues or indicators that affect your trust in a matrimony profile?
    \item Have you or anyone you know encountered negative situations like impersonation, where others pretended to be them on the internet?
    \item What warning signs or red flags do you watch out for to determine whether a matrimony profile is legitimate?
    \item What kind of responses have you received to your profile so far?
\end{enumerate}
    \textbf{Admin’s responsibilities}
\begin{enumerate}
    \item What role do admins have in providing a safe space for sharing personal information?
    \item How do admins maintain privacy and security in Facebook matrimony groups?
    \item How should admins communicate regarding privacy risks?
    \item What are they currently doing / not doing for privacy?
    \item How can the admin make these groups a safer place for everyone?
    \item Have you had any privacy or security issues, and how did you resolve them?
\end{enumerate}
   \textbf{Suggestions for platforms}  
\begin{enumerate}
   \item How can platforms like Facebook better support user’s goals and privacy?
    \item How do you think Facebook can support protecting user’s privacy?
    \item Do you have any suggestions for increasing security and privacy in Facebook matrimony groups? 
\end{enumerate}

\newpage

\section{Matrimonial Profile Template \& Samples}
\label{tab:profile template, samples}

\begin{figure}[htbp]
    \centering
    \includegraphics[width=\linewidth]{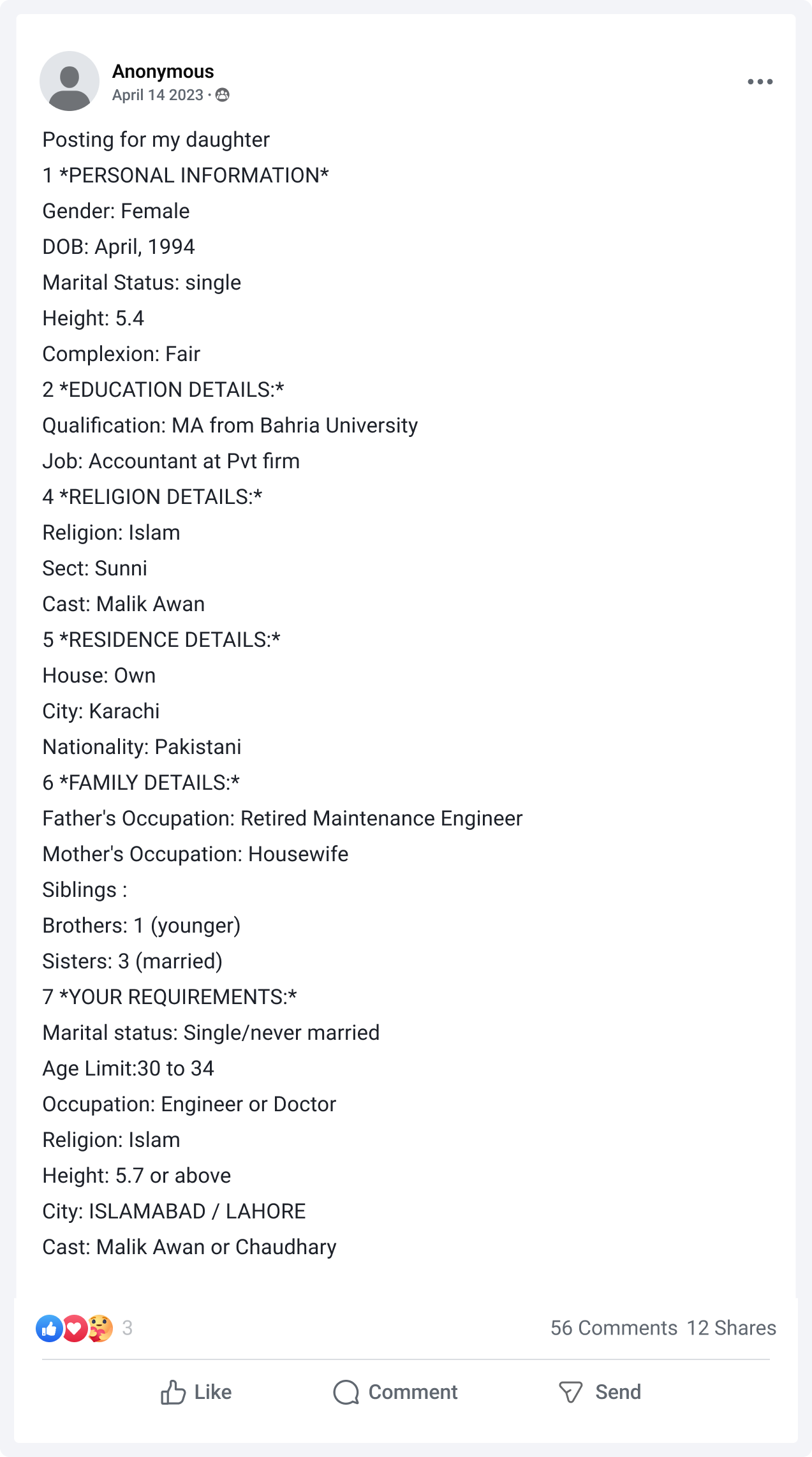}
    \caption{Facebook matrimonial anonymous post (mockup).}
    \label{fig:anonymous}
\end{figure}


\begin{figure}[htbp]
    \centering
    \includegraphics[width=\linewidth]{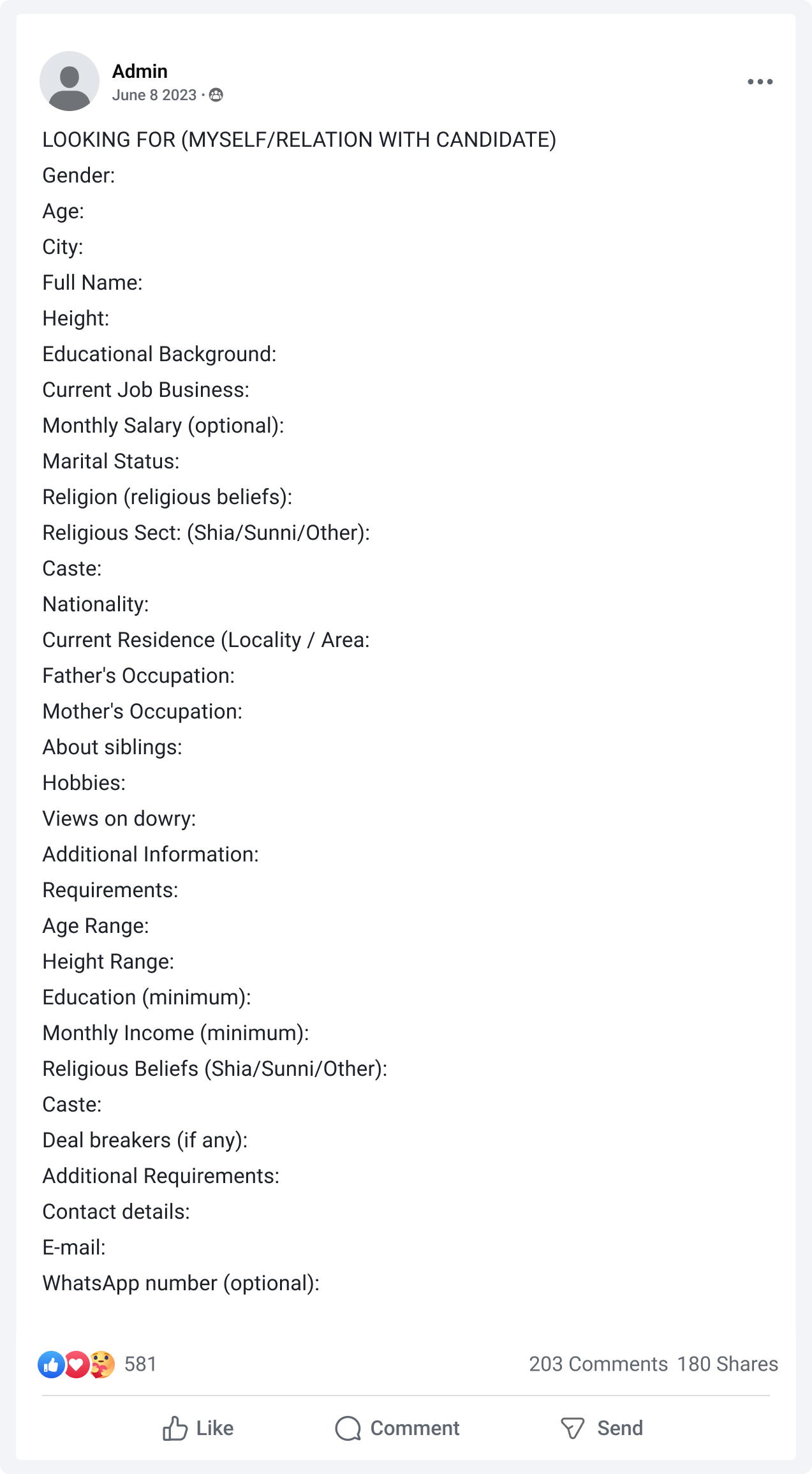}
    \caption{Matrimonial profile template (mockup).}
    \label{fig:template}
\end{figure}

\clearpage

\onecolumn

\section{Codebook}
\label{tab:codebook}
\footnotesize
\begin{longtable}
{p{4.5cm} p{5.5cm} p{7.5cm}}
\caption{Interview codebook.} \label{Interview data codebook} \\
  \toprule
  \textbf{Category} & \textbf{Codes} & \textbf{Definition or example} \\ \midrule \endfirsthead
  \toprule
  \textbf{Category} & \textbf{Codes} & \textbf{Definition or example} \\ \midrule \endhead
\rowcolors{1}{white}{gray!10}
Demographics questions & Member of matrimonial group & Participants answer if they are part of matrimonial groups (Yes/No) \\
& Name of platforms/apps used & Participants mention different applications/platforms they use to find matches. Example, WhatsApp, Facebook, etc \\
& Group name on Facebook & Participants mentioned a few matrimonial group names on Facebook that they are part of. Examples, Two Rings, Soul sister, etc. \\
& Reasons for using Facebook & Motivations behind choosing Facebook for finding matches. Examples, security provided, issues with other platforms, friends/family recommendations.\\
& Introduction to groups & How participants learned about matrimonial groups. Example, through family, friends, social media, etc.\\
& Posted profile on group & Participants mentioned multiple ways of posting details on matrimonial groups. For self, friends/family, anonymously, or never posted. \\\hline
Type of information can be posted or kept private & Can be posted & details participants mentioned that they add on posts. Example, picture, contact, gender, religion, age, caste, interests, languages spoken, family details, height, name, education, job details, etc.\\
& Kept private & details participants mentioned that they don’t prefer to add on posts. Example, full name includes family name, property or wealth related details, salary or income, skin complexion, email address, contact number, pictures, etc\\ \hline
Perceptions of Sharing Information Online & Online harassment & Fear of online bullying, threat, harassment. \\
& Personal Information &fear of sharing personal information due to safety and privacy concerns. \\
& Sensitive topic & they feel it's a sensitive topic to look for proposals online. \\
& No risks & they do not perceive any risks. \\ \hline
Privacy and Security incidents &  Others' experiences & privacy, security issues heard from others. \\
& Self experiences & own experiences of privacy, security issues. \\ \hline
Successful and unsuccessful match experiences & Success stories - others & reports of successful matches from others. \\
& Success stories - personal & participants’ own experience of finding successful matches on matrimonial groups. \\
& Unsuccessful stories - others & reports of rejections, non-compatible experiences shared by others. \\
& Unsuccessful stories - personal & participants’ own experience of rejections, non-compatible experiences on matrimonial groups.\\ \hline
Trust in Profiles & Anonymous posts & a few participants chose not to trust posts from anonymous users, and they trust posts where their Facebook profiles are tagged or mentioned.\\
& Trust based on gender & they trust the posts based on the gender of the user who posts on the group.\\
& Profile transparency & participants trust profiles that provide clear and detailed information, such as activity history and profile completeness. \\
& Premium groups are serious &  the matrimonial groups, where users pay to be part of, have more serious people who intend to find matches. \\ \hline
Trust in Matches & Online and offline verification & using both digital and real-world sources to validate information provided by the match, for example, neighbours, social circles, workplace, social media footprint, cross-checking details on the proposal, etc.\\
& Involvement of friends/family & trusting the match based on family and friends' involvement. \\
& Contact/meet in-person & contact the match or meet them in-person, and get information. \\ \hline
Safety Measures & Limit information shared & participants choose to avoid sensitive details, for example, avoiding full name, contact number, income, or exact location.\\ 
& Avoid/report suspicious profiles & participants prefer to ignore or report suspicious profiles.\\ \hline
Admin Responsibilities & Match profiles & helping members of the group to find a suitable match. \\
& Provide better privacy settings & safeguard the group with safety settings, as they have permissions. \\
& Provide members with security and privacy tips & educating users on the groups with important safety practices. \\
& Verification of user accounts and posts & go through members’ profiles and check if the details provided in the profile and posts are genuine. \\
& Restrict negative information & admins have to refrain from posting negative or provoking information on the groups, and also take measures to prevent members from posting negative information. \\
& Limited authority &  the admin will not have enough resources to help the members in some cases.\\ \hline
Reasons for family's approval and disapproval & To approve & the reasons for which their family agreed to use matrimonial groups to find matches. For example, they trust the online groups, they left the decision to the person who is finding a match, didn’t have any opinion, etc.\\
& To disapprove & the reasons for which their family didn’t want to use online groups for match finding. Example, didn’t know how online matchmaking works, no resources like mobiles, privacy concerns, etc. \\ \hline
Factors involved in approving matches & Good family and education & the family background, and the education of the match are important.\\
& Family approval & before finalizing a match found online, their family’s approval is important. \\ \hline
Reasons to use matrimonial groups & International match finders & the groups are beneficial for people who stay abroad and find matches in their home country.\\
& Will know current trends & being part of groups can give us information about what people are interested in and how matchmaking happens online.\\
& No charges (from Rishta aunties) & unlike offline people who charge money to find matches, few online groups don’t charge money.\\
& Finding a match & the groups are very helpful to find suitable matches. \\ 
& Admin validation of profiles & sometimes admins verify the profiles of the members in the group, and it helps.\\
& To find matches in older age & helps to find matches in any age range. \\
& Couldn't find matches offline & helps users who couldn’t find a suitable match offline. \\
& Many people reach out & on the groups many people reach out intending to find a match. \\ \hline
Threats associated with matrimonial groups & Derogatory comments & receiving negative or bullying comments on posts based on physical aspects, etc.\\
& Threats &  receiving threats by trapping users.\\
& Fake accounts & not being able to validate fake accounts and posts.\\
& Offline match makers are better & compared to the cons of using groups participants felt the offline process is better. \\ 
& No accountability or correct information & there is no one responsible or helpful in the group to validate information.\\
& Posts pictures & posting pictures of the person exploiting their privacy.\\
& Relatives/Friends knowing & relatives and others knowing that one is searching for a match online, which is considered demeaning sometimes. \\ 
& Misuse of personal information & some of them misuse the information on the groups with wrong intentions, and it is considered unsafe to share sometimes.\\ \hline
Alternate options to find matches & Dating apps & dating apps can be used to find matches.\\
& Matrimonial websites & a few premium groups that charge money to be part of can be used.\\
& Social circle/relatives & people from social circles, friends, and family also help to find a suitable match.\\
& Rishta aunties & women who do match-making as a profession and charge money for their services.\\  \hline
Suggestions for platforms & Protect from download option &pictures posted on groups should be protected from download option.\\
& Protect from screenshots & pictures posted on groups should get protected from the screenshot option.\\
& Mitigate data misuse & platforms should implement measures to protect users' information.\\
& Verify accounts & platforms should actively verify the accounts on their platforms.\\ 
& Better reporting mechanisms & platforms should have good reporting mechanisms for users to report issues.\\
& Detect fake accounts & platforms should identify the fake accounts and inform users to be aware of them.\\
& Ensure participants’ accountability & platforms should implement measures to hold participants accountable for their actions and make sure they face consequences for misinformation, harassment, violating group policies, etc. \\ 
& No suggestions & some participants believe that platforms already have sufficient features and do not see a need for further improvements.\\ \hline
Suggestions from participants & Approach admin & contact admin with issues on the groups.\\
& Approach cybersecurity team & participants report safety issues faced on platforms to the government agencies like the cybersecurity team.\\
& Block harassment & participants block and report users who bully, harass etc.\\ \hline
Types of people on the platform & Fake posts & people who post fake details on groups for fun or other intentions.\\
& Sabotages posts & people who leave false comments or accusations on others' posts.\\
& Curious ones & people who are just curious to know whats happening on matrimonial groups.\\ \hline
Misc statements & & Few miscellaneous(outstanding) statements provided by participants in interviews.\\
\bottomrule
\end{longtable}

\end{document}